\documentclass[preprint,12pt]{elsarticle}

\usepackage{amssymb}
\usepackage{amsmath}
\usepackage{multicol}
\usepackage{natbib}
\usepackage[version=4]{mhchem}
\usepackage{color}
\usepackage{soul}
\usepackage{hyperref}

\newlength{\onecol}
\newlength{\twocol}
\newcommand{\ipi}{{i-PI}}
\newcounter{bla}

\newcommand{\rev}[1]{#1}

\newcommand{\tm}{\text{m}} \newcommand{\tf}{\text{f}}

\newcommand{\bq}{\mathbf{q}}

\journal{Computer Physics Communications}

\begin{document}

\begin{frontmatter}

\title{ \ipi{} 2.0: A Universal Force Engine for Advanced Molecular Simulations}

\author[a]{{ Venkat Kapil}}
\author[b]{{ Mariana Rossi}}
\author[d,n]{{ Ondrej Marsalek}}
\author[a]{{ Riccardo Petraglia}}
\author[b]{{ Yair Litman}}
\author[f]{{ Thomas Spura}}
\author[a]{{ Bingqing Cheng}}
\author[a]{{ Alice Cuzzocrea}}
\author[a]{{ Robert H. Mei{\ss}ner}}
\author[a]{{ David M. Wilkins}}
\author[a]{{ Przemys\l{}aw Juda}}
\author[a]{{ S\'ebastien P. Bienvenue}}
\author[j]{{ Wei Fang}}
\author[f]{{ Jan Kessler}}
\author[c]{{ Igor Poltavsky}}
\author[k]{{ Steven Vandenbrande}}
\author[k]{{ Jelle Wieme}}
\author[m]{{ Clemence Corminboeuf}}
\author[f]{{ Thomas D. K\"uhne}}
\author[l]{{ David E. Manolopoulos}}
\author[d]{{ Thomas E. Markland}}
\author[j]{{ Jeremy O. Richardson}}
\author[c]{{ Alexandre Tkatchenko}}
\author[e]{{ Gareth A. Tribello}}
\author[k]{{ Veronique Van Speybroeck}}
\author[a]{{ Michele Ceriotti\corref{author}}}

\cortext[author] {Corresponding author.\\\textit{E-mail address:} michele.ceriotti@epfl.ch}
\address[a]{Laboratory of Computational Science and Modeling, Institute of Materials, {\'E}cole Polytechnique F{\'e}d{\'e}rale de Lausanne, 1015 Lausanne, Switzerland}
\address[b]{Fritz Haber Institute of the Max Planck Society, Theory Department, Faradayweg 4-6, 14195 Berlin, Germany}
\address[c]{Physics and Materials Science Research Unit, University of Luxembourg, L-1511 Luxembourg}
\address[d]{Department of Chemistry, Stanford University, Stanford, California, 94305, USA}
\address[e]{Atomistic Simulation Centre, School of Mathematics and Physics, Queen's University Belfast, Belfast, BT7 1NN, United Kingdom}
\address[f]{Department of Chemistry and Center for Sustainable Systems Design, University of Paderborn, Warburger Str. 100, D-33098 Paderborn, Germany}
\address[j]{Laboratory of Physical Chemistry, ETH Zurich, 8093 Zurich, Switzerland}
\address[k]{Center for Molecular Modeling, Ghent University, 9052 Zwijnaarde, Belgium}
\address[l]{Physical and Theoretical Chemistry Laboratory, South Parks Road, Oxford OX1 3QZ, UK}
\address[m]{Laboratory for Computational Molecular Design, Institute of Chemical Sciences and Engineering, {\'E}cole Polytechnique F{\'e}d{\'e}rale de Lausanne, 1015 Lausanne, Switzerland}
\address[n]{Charles University, Faculty of Mathematics and Physics, Ke Karlovu 3, 121 16 Prague 2, Czech Republic}

\begin{abstract}

Progress in the atomic-scale modelling of matter over the past decade has been tremendous. This progress has been brought about by improvements in methods for evaluating interatomic forces that work by either solving the electronic structure problem explicitly, or by computing accurate approximations of the solution and by the development of techniques that use the Born-Oppenheimer (BO) forces to move the atoms on the BO potential energy surface. 
As a consequence of these developments it is now possible to identify stable or metastable states, to sample configurations consistent with the appropriate thermodynamic ensemble, and to  estimate the kinetics of reactions and phase transitions.  
All too often, however, progress is slowed down by the bottleneck associated with implementing new optimization algorithms and/or sampling techniques into the many existing electronic-structure and empirical-potential codes.
To address this problem, we are thus releasing a new version of the \ipi{} software.  This piece of software is an easily extensible framework for implementing advanced atomistic simulation techniques using interatomic potentials and forces calculated by an external driver code. 
While the original version of the code\cite{ceri+14cpc} was developed with a focus on path integral molecular dynamics techniques, this second release of \ipi{} not only includes several new advanced path integral methods, but also offers other classes of algorithms.  In other words, \ipi{} is moving towards becoming a universal force engine that is both modular and tightly coupled to the driver codes that evaluate the potential energy surface and its derivatives. 
\end{abstract}


\end{frontmatter}

{\bf PROGRAM SUMMARY}

\begin{small}
\noindent
{\em Manuscript Title: } \ipi{} 2.0: A Universal Force Engine for Advanced Molecular Simulations \\
{\em Authors: } V. Kapil, M. Rossi, O. Marsalek, R. Petraglia, Y. Litman, T. Spura, B. Cheng, A. Cuzzocrea, R. Mei{\ss}ner, D. Wilkins, P. Juda, S. Bienvenue, J. Kessler, I. Poltavsky, S. Vandenbrande, J. Wieme, C. Corminboeuf, T. D. K\"uhne, D. Manolopoulos, T. Markland, J. Richardson, A. Tkatchenko, G. Tribello, V. Van, M. Ceriotti \\
{\em Program Title: } \ipi{} \\
{\em Journal Reference:}                                      \\
{\em Catalogue identifier:}                                   \\
{\em Licensing provisions: GPLv3, MIT}                                   \\
{\em Programming language: } Python                                   \\
{\em Computer: } multiple architectures                                                \\
{\em Operating system: } Linux, Mac OSX, Windows                                       \\
{\em RAM: } less than 256 Mb                                          
{\em Classification:} 7.7                                        \\
{\em External routines/libraries:}  NumPy                       \\
{\em Nature of problem:}\\
Lowering the implementation barrier to bring state-of-the-art sampling and atomistic modelling techniques to \emph{ab initio} and empirical potentials programs
   \\
{\em Solution method:}\\
Advanced sampling methods, including path-integral molecular dynamics techniques, are implemented in a Python interface. Any electronic structure code can be patched to receive the atomic coordinates from the Python interface, and to return the forces and energy that are used to integrate the equations of motion, optimize atomic geometries, etc. 
   \\
{\em Restrictions:}\\
This code does not compute interatomic potentials, although the distribution includes sample driver codes that can be used to test different techniques using a few simple model force fields. 
   \\
{\em Running time:}\\
   Depends dramatically on the nature of the simulation being performed. A few minutes 
for short tests with empirical force fields, up to several weeks for production calculations
with \emph{ab initio} forces.
 \\

\end{small}

\section{Introduction}

Atomic-scale modelling is constantly improving in terms of the efficiency and accuracy of its predictions. 
This is due to a fortunate combination of the baseline increase of available computational power, the availability of better electronic-structure methods and their (semi-)empirical approximations, and the development of sampling schemes that can efficiently map the static and dynamic properties of matter based on the underlying description of the interatomic potential energy surface. 
In such a rapidly evolving and interdisciplinary field of research it is important that complex and/or newly-developed schemes become available to end-users with the minimum possible implementation overhead. 
With this goal in mind, a few years ago the first version of \ipi{} was released.  This code was designed as an easily-extensible Python package for performing molecular dynamics and path integral molecular dynamics simulations~\cite{ceri+14cpc}. 
By using a socket interface to request and gather interatomic forces and potential energies from an external driver code, \ipi{} provided a flexible infrastructure that could be connected to different types of electronic structure and empirical force field packages.
This code therefore provided a platform that allowed researchers in the chemistry and materials science communities to easily incorporate state-of-the-art methods for dealing with the quantum nature of atomic nuclei in their simulations~\cite{mark-ceri18nrc}.

In the years after the first release of \ipi{}, several new methods for accelerating the simulations of quantum nuclei have been developed. In addition, the user base of \ipi{} has grown considerably, and many researchers have used the code to implement advanced modelling techniques that go beyond (path integral) molecular dynamics. 
This paper presents a second release of \ipi{} that includes these new developments and a refactored core structure that is more suited for the implementation of, amongst other things, techniques that involve multiple replicas of a physical system. 
After giving an overview of the rationale underlying the updated infrastructure and of the new features that have been included in this release, a few examples of some of the most recently-implemented functionalities will be presented in order to showcase the power of \ipi{} as a universal force engine for molecular simulations.

\section{Program overview}

\begin{figure}
\centering\includegraphics[width=0.5\columnwidth]{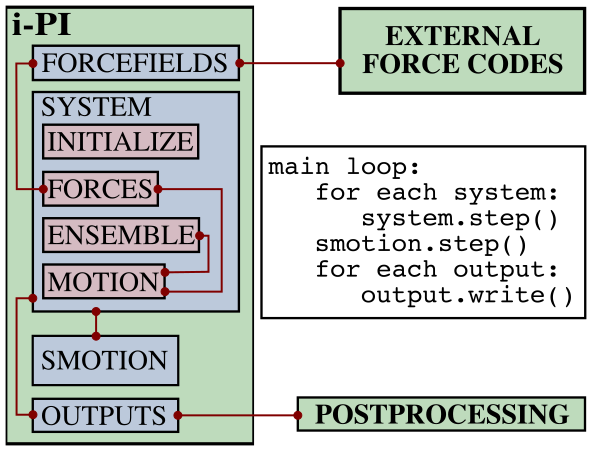}
\caption{
A schematic representation of the structure of \ipi{}. \rev{Green boxes identify programs, red and blue boxes identify classes, and lines represent information exchange between classes, all in a loose sense. This structure is reflected in the XML input format of i-PI. 
The {\bf System} (or a collection of systems) is evolved in steps, which are compounded by collective moves that can exchange information between systems ({\bf SMotion}) and by an {\bf Output} management. 
The i-PI core is complemented by external drivers (``force codes") that compute energy and forces -- the interface with which is managed by a {\bf ForceField} class -- and by post-processing tools that can compute observables from the output of i-PI. }
\label{fig:structure}}
\end{figure}

In keeping with the original goals set out for the first version of \ipi{}\cite{ceri+14cpc}, this new release is still built around the fact that force evaluation is the bottleneck of most materials modelling simulations. 
For this reason, even though several performance improvements have been implemented, in particular to accelerate I/O, the main infrastructural changes are motivated by a recognition that the code needed a clearer structure that better reflects the basic functions of a force engine. 
As shown in Figure~\ref{fig:structure}, the core element of \ipi{} is a {\bf System} class. This class provides a description of the physical system under study by explaining how it should be initialized, how energy and forces should be computed and what statistical ensemble should be sampled. If the system class is set up correctly, one can then use a 
{\bf Motion} class to describe how atoms should be evolved in steps, either through molecular dynamics, Monte Carlo or to minimize (or maximize) energy, or simply to compute functions or derivatives of the energy or the force.
Multiple {\bf System} classes can be defined and evolved in parallel, and one can even define evolution steps that combine different systems through a {\bf SMotion} (Systems Motion) class -- such as for instance exchanges between parallel replicas. 
As in the first release, a series of {\bf Forcefield} classes can be defined that specify how the forces that drive the different systems should be computed. This procedure will possibly take place with the help of external driver codes that connect to \ipi{} through a socket interface. 
\rev{As explained in more detail in the paper accompanying the first release of \ipi{}~\cite{ceri+14cpc}, there is a conceptual difference between {\bf Forcefield} classes that describe how to compute energy and forces given a set of atomic coordinates, and the {\bf Forces} class that describes how the (potentially) many potential components should be combined to describe in full the energetics of the system. 
As we discuss below, the possibility of defining multiple force components simplifies greatly the setup of complex simulation schemes, such as those involving multiple time stepping in real or imaginary time.
In this version we also separated the definition of the thermodynamic boundary conditions -- that were previously part of the parameters of the molecular dynamics integrator -- into an {\bf Ensemble} class. As discussed in Section~\ref{sec:remd}, this change in architecture simplifies greatly the implementation of advanced sampling schemes based on replica exchange. 
}
Finally, outputs can be generated in a customisable form, and then further analyzed by a series of post-processing tools. 

\section{Program features}

Over the past two years, \ipi{} has accumulated a large number of new features, that go well beyond its original purpose as a code for performing path integral molecular dynamics simulations. Many people have been involved in designing, executing and funding the development of these features, and their involvement has been acknowledged by making them part of the author team for this publication. 
Here, the main features that are available in this new release of \ipi{} are listed.  The main material contributors to each feature are indicated together with a reference to the papers describing the theory and the implementation in i-PI. In addition to the implementation of these simulation methods, considerable effort has been devoted to optimizing, documenting and streamlining the code and the overall infrastructure.  The contributions of R. Petraglia, V. Kapil, M. Rossi, O. Marsalek, T. Spura, J. Przemyslaw and M. Ceriotti have been particularly significant in these regards. 

The features in the original version of the code \cite{ceri+14cpc} are still available:
\begin{itemize}
\item molecular dynamics and PIMD in the {\em NVE}, {\em NVT} and {\em NPT} ensembles, with the high-frequency internal vibrations of the path propagated in the normal-mode representation~\cite{ceri+10jcp} \rev{to allow for longer time steps};
\item ring polymer contraction~\cite{mark-mano08jcp,mark-mano08cpl}, implemented 
by exposing multiple socket interfaces to deal with short and long-range components of the potential energy separately; \rev{treating different components that have different computational cost and characteristic time scale separately can reduce substantially the overall effort associated with a simulation}
\item efficient stochastic velocity rescaling \cite{buss+07jcp} and 
path integral Langevin equation thermostats~\cite{ceri+10jcp} \rev{to sample efficiently the canonical ensemble}; 
\item various generalized Langevin equation (GLE) thermostats,
including the optimal sampling~\cite{ceri+09prl,ceri+10jctc}, quantum 
~\cite{ceri+09prl2}, and $\delta$~\cite{ceri-parr10pcs} thermostats; the parameters 
\rev{for different GLE flavors and the conditions in which they should be applied can be obtained from a separate website~\cite{gle4md};}
\item mixed path integral--generalized Langevin equation techniques for 
accelerated convergence, including both PI+GLE~\cite{ceri+11jcp} and the 
more recent and effective version PIGLET~\cite{ceri-mano12prl}; \rev{these techniques reduce the number of path integral replicas needed, while allowing for systematic convergence;}
\item all the standard estimators for structural properties, the quantum 
kinetic energy, pressure, etc.;
\item more sophisticated estimators such as the scaled-coordinate 
heat capacity estimator~\cite{yama05jcp}, 
estimators to obtain isotope fractionation free energies by re-weighting a 
simulation of the most abundant isotope~\cite{ceri-mark13jcp}, and 
a displaced-path estimator for the particle momentum distribution~\cite{lin+10prl};
\item the infrastructure that is needed to perform approximate quantum dynamics calculations such as ring polymer molecular dynamics (RPMD)~\cite{crai-mano04jcp,habe+13arpc} and centroid molecular dynamics
(CMD)~\cite{cao-voth93jcp,cao-voth94jcp}.
\end{itemize}

To these, several additional methods have been made available, including several last-generation path-integral techniques:
\begin{itemize}
\item reweighted fourth-order path integral MD (M. Ceriotti, G.A.R. Brain)~\cite{ceri+12prsa,jang-voth01jcp}; \rev{this method makes it possible to obtain fourth-order statistics by re-weighting second-order trajectories; attention should be paid to avoid statistical inefficiencies;}
\item finite-differences implementation of fourth-order path integrals (V. Kapil, M. Ceriotti)~\cite{kapi+16jcp2}; \rev{this schemes enables explicit fourth-order path integral simulations, that converge faster than conventional Trotter methods};
\item perturbed path integrals (I. Poltavsky)~\cite{polt-tkat16cs}; \rev{essentially, a truncated cumulant expansion of fourth-order reweighting, that often enables fast convergence avoiding statistical instability; }
\item open path integrals and momentum distribution estimators (V. Kapil, A. Cuzzocrea, M. Ceriotti) \cite{kapi+18jpcb}; \rev{makes it possible to compute the particle momentum distribution including quantum fluctuations of nuclei;} 
\item quantum alchemical transformations (B. Cheng) \cite{liu+13jpcc,chen+16jpcl}; \rev{Monte Carlo exchanges between isotopes of different mass, useful to sample isotope propensity for different molecules or environments;}
\item direct isotope fractionation estimators (B. Cheng, M. Ceriotti) ~\cite{chen-ceri14jcp}; \rev{avoid thermodynamic integration to obtain isotope fractionation ratios;}
\item spatially localized ring polymer contraction (M. Rossi, M. Ceriotti) \cite{litm+17jcp}; \rev{simple contraction scheme for weakly bound molecules, e.g. on surfaces;}
\item ring polymer instantons (Y. Litman, J.O. Richardson, M. Rossi) \cite{litm-preparation}; \rev{evaluation of reaction rates and tunnelling splittings for molecular rearrangements and chemical reactions;}
\item thermodynamic integration (M. Rossi, M. Ceriotti) \cite{ross+16prl}; \rev{classical scheme to compute free energy differences;}
\item geometry optimizers for minimization and saddle point search (B. Helfrecht, R. Petraglia, Y. Litman, M. Rossi) \cite{ross+16prl}
\item harmonic vibrations through finite differences (V. Kapil, S. Bienvenue)  \cite{ross+16prl}; \rev{simple evaluation of the harmonic Hessian;}
\item multiple time stepping (V. Kapil, M. Ceriotti) \cite{kapi+16jcp}; \rev{accelerated simulations by separating slow and fast degrees of freedom into different components of the potential energy;}
\item metadynamics through a PLUMED interface (G. Tribello, M. Ceriotti); \rev{simulation of rare events and free energy calculations;}
\item replica exchange MD (R. Petraglia, R. Meissner, M. Ceriotti)~\cite{petr+15jcc}; \rev{accelerated convergence of averages by performing Monte Carlo exchanges of configurations between parallel calculations}
\item thermostatted RPMD~\cite{ross+14jcp}, including optimized-GLE TRPMD~\cite{ross+18jcp}; \rev{reduces well-known artifacts in the simulation of dynamical properties by path integral methods;}
\item dynamical corrections to Langevin trajectories (M. Rossi, V. Kapil, M. Ceriotti)~\cite{ross+18jcp}; \rev{eliminates the artifacts introduced into dynamical properties by the presence of thermostats;}
\item fast forward Langevin thermostat (M. Hijazi, D. M. Wilkins, M. Ceriotti); \rev{a simple scheme to reduce the impact of strongly-damped Langevin thermostats on sampling efficiency;} \cite{hija+18jcp}
\item Langevin sampling for noisy and/or dissipative forces (J. Kessler, T. D. K\"uhne)~\cite{PhysRevB.73.041105,PhysRevLett.98.066401}; \rev{suitable to stabilize and correct the artifacts that are introduced in MD trajectories by different extrapolation schemes;}

\end{itemize}

\section{Examples of New Features}

In order to demonstrate the flexibility of the \ipi{} software, a few examples of selected features that have been made available in this release are discussed in the following section.

\subsection{Replica exchange molecular dynamics}\label{sec:remd}

Replica exchange refers to a large family of accelerated sampling methods in which multiple simulations (e.g.\,MD) are performed in parallel for the same system, with each simulation sampling from a different ensemble. Many different variations on this theme have been proposed, including running parallel simulations at different temperatures~\cite{Sugita1999} or pressures~\cite{Okabe2001}, using different bias potentials on different replicas ~\cite{pian-laio07jpcb} or having replicas that experience different Hamiltonians because various parts of the interatomic potential have been scaled differently~\cite{Sugita2000}. 
In all cases, the ensemble-preserving parallel evolution of the different systems is combined with Monte Carlo moves that exchange the configurations between pairs of replicas.

To understand the general idea -- and the way it has been implemented in this version of i-PI --  consider that each ensemble $\aleph$ effectively describes a probability distribution for the positions and momenta of a system, $P_\aleph(\mathbf{p},\mathbf{q})$. 
If two of these distributions, $P_\aleph$ and $P_\beth$, are selected at random, one can build Monte Carlo moves that preserve the distribution of both ensembles by performing a swap and accepting or rejecting the move with a Metropolis rate     
\begin{equation}
a_{1\leftrightarrow 2} = 
\frac{P_\aleph(\mathbf{p}_1,\mathbf{q}_1) P_\beth(\mathbf{p}_2,\mathbf{q}_2)}
{P_\beth(\mathbf{p}_1,\mathbf{q}_1) P_\aleph(\mathbf{p}_2,\mathbf{q}_2)}.
\label{eq:remd-accept}
\end{equation}
where $\left\{\mathbf{p}_1,\mathbf{q}_1\right\}$ and $\left\{\mathbf{p}_2,\mathbf{q}_2\right\}$ are configurations that are initially distributed according to $P_\aleph$ and $P_\beth$ respectively. 
Notice that only the ratios between the probability distributions need to be computed. Furthermore, in all cases, these can be cast in terms of the exponential of a difference between (generalized) Hamiltonians computed with different parameters. 

In practice, we implemented replica exchange in \ipi{} by performing a swap between the ensemble definitions. 
All the information on temperature, pressure, bias and the scaling factors for different components of the potential is contained in the {\bf Ensemble} class. Therefore, it suffices to swap only the values of all these parameters as the \emph{depend} object machinery will then update all the necessary derived quantities (including e.g.\,the spring constant in a PIMD simulation), so that the probability factors after the trial swap can be computed, and used to decide on the acceptance. 
Whenever it is possible the momenta are rescaled in a reversible way after a swap and before computing Eq.~\eqref{eq:remd-accept}. This eliminates the dependence of the probability ratios on the values of momenta and also increases the likelihood for acceptance. If a move is rejected, both ensembles are restored to their initial state.

\begin{figure}[htpb]
\includegraphics[width=\textwidth]{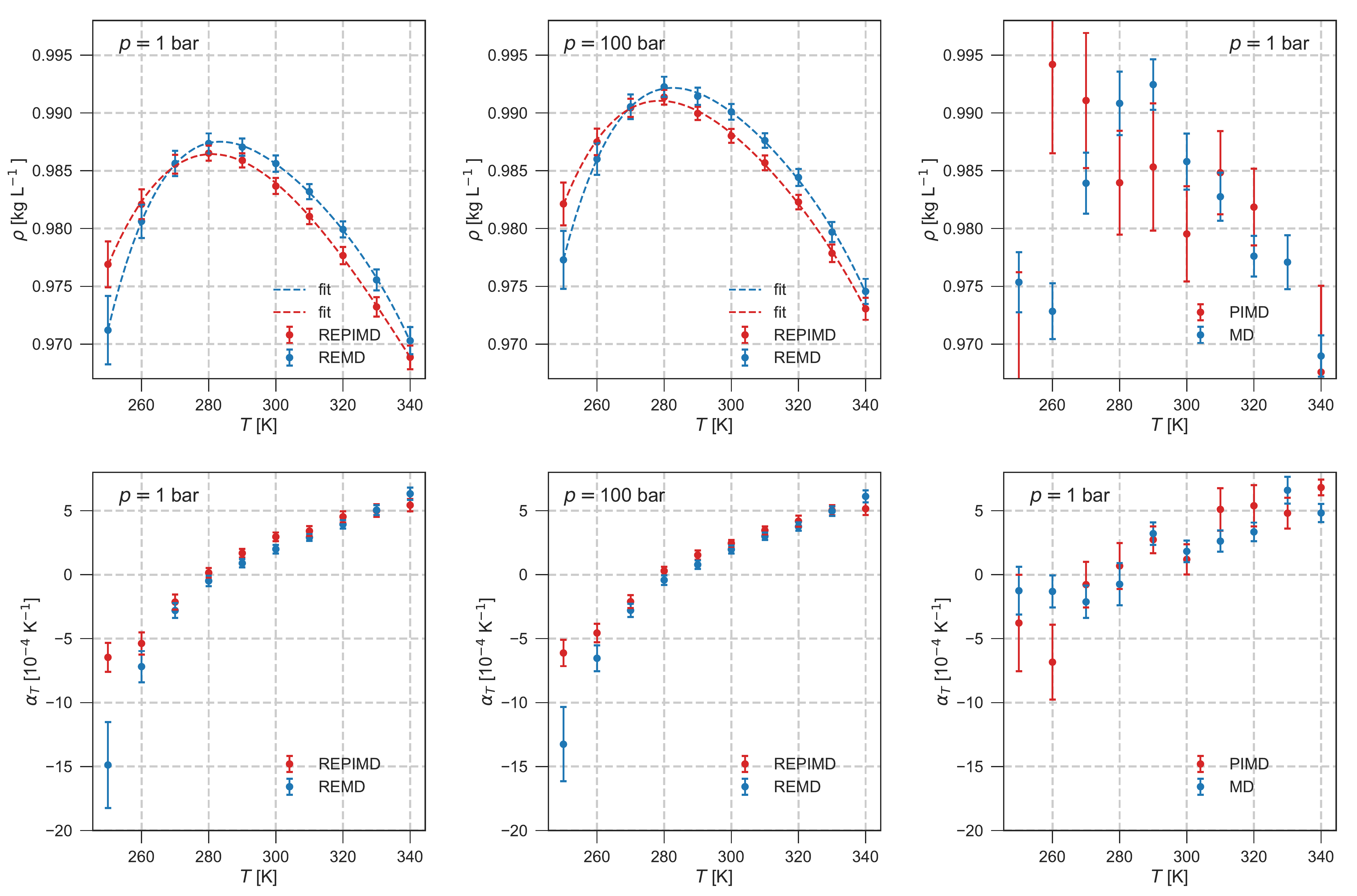}
\caption{
 The panels show the temperature dependence for the density (left) and the coefficient of thermal expansion (right) calculated for 64 molecules of q-TIP4P/f water using path integral (red) and classical MD (blue). While panels (a) and (b) were calculated at 1 bar, (c) and (d) were calculated at 100 bar from $NpT$ replica exchange simulations. Panels (e) and (f) were obtained at 1 bar without any replica exchange moves. The dotted lines on the upper left and upper middle panel are fourth degree polynomials fitted to the averages obtained from the simulations.  All averages are calculated from 500 ps simulations. 
\label{fig:remd_density}
}
\end{figure}

The i-PI implementation of replica exchange has already been used to sample organic chemical reactions and conformational transitions with density functional tight binding~\cite{petr+15jcc}. Here, we provide a simple demonstration that illustrates the power of this approach using both classical and path integral REMD. Figure \ref{fig:remd_density} shows the calculated density and the coefficient of thermal expansion as a function of temperature (250 K to 340 K in steps of 10K) at 1 bar and 100 bar for 64 molecules of q-TIP4P/f water \cite{habe+09jcp}. The energies and forces were evaluated using LAMMPS~\cite{plim95jcp}. 
A total of 20 $NpT$ ensembles, one for each tuple of pressure and temperature,  were simulated using the BZP barostat~\cite{buss+09jcp}
with a time constant of 1200 fs. The barostat was coupled to a white noise Langevin thermostat with a time constant of 100 fs. The physical degrees of freedom were attached to a PILE-G~\cite{ceri+10jcp} and stochastic velocity rescaling thermostat~\cite{buss+07jcp} with a time constants of 100 fs, respectively for PIMD and MD. The equations of motion of the system were integrated using a conservative time step of 0.25 fs. 
Exchanges between the ensembles were attempted every 10 steps for classical simulations and every step for quantum simulations. To compare the efficiency of the replica exchange scheme with respect to straightforward molecular dynamics, several independent (path integral) molecular dynamics runs were also performed at the 20 thermodynamic conditions, without including swaps. 
Figure \ref{fig:remd_density} compares replica exchange (panels (a) - (d)) and conventional (PI)MD (panels (e) and (f)) for the calculation of the density versus temperature profile and the calculation of the coefficient of thermal expansion.
REMD reaches satisfactory convergence with trajectories that involve fewer than 500 ps of dynamics so the change in the density profile as the pressure is increased from 1bar to 100 bar can be easily observed from the converged results. Plain (PI)MD would require tens of ns \cite{habe-mano09jcp} to converge these observables to the accuracy of replica exchange simulations. The curves also allow one to appreciate the subtle shift in the temperature of maximum density and the coefficient of thermal expansion at low temperatures towards the experimental result which occurs when nuclear quantum effects are switched on.

\begin{figure}[tbpt]
\centering\includegraphics[width= \columnwidth]{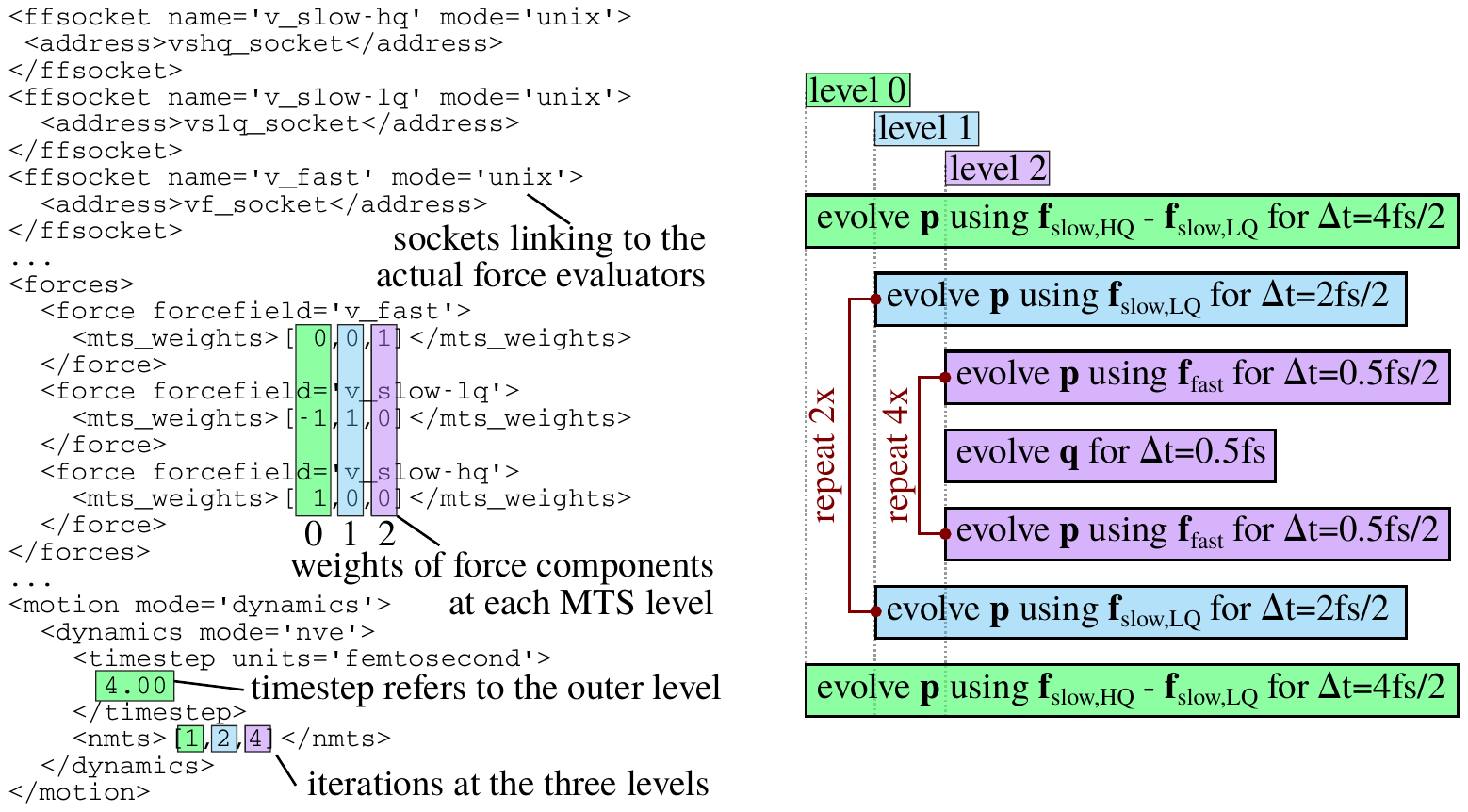}
\caption{ 
A commented snipped of an \ipi{} input file containing the essential components of a MTS simulation. The right portion of the figure schematically represents the steps in an integrator that correspond to such input. 
\label{fig:mts-scheme}}
\end{figure}

\rev{
\subsection{Multiple time step integrators}

Multiple time step (MTS) integration has been implemented in i-PI in Ref.~\citenum{kapi+16jcp}, and can be used whenever the potential energy of the system can be split into a slowly-varying and a quickly-varying part, $V=V_\text{slow}+V_\text{fast}$. 
The implementation is extremely flexible, and allows for arbitrarily complex setups in which many levels of time stepping are combined. Consider for example a situation in which the potential is split in a slow and a fast part, and in addition the slow potential can be computed using an accurate but demanding method, that yields $V_\text{slow,HQ}$, or an approximate, inexpensive scheme, $V_\text{slow,LQ}$. The difference between the two terms is small and varies even more slowly.
An \ipi{} input that realizes this complex simulation set up is shown in Fig.~\ref{fig:mts-scheme}. Several levels of MTS can be defined in the {\bf Motion} section, and the time step corresponds to the outer level. The time steps corresponding to the different MTS levels are controlled by the \texttt{nmts} vector. Entries in this vector corresponds to the number of times that the propagation at a certain level should be repeated before doing one time propagation step in the level directly above. The outermost level corresponds to an actual full step of the simulation. Outputs can only be generated at this level, and the time step indicated in the input refers to this outermost level.

The forces that are active at each MTS level are obtained by combining multiple components, based on weights that are defined for each level and each component. 
For instance, if one wants to have a single component active at a desired level, its {\texttt{mts\_weights}} vector should contain only zeroes, except for a ``1'' at the desired level. The difference between two components can be realized by using negative and positive weights for different components at the same MTS level. 
The automatic dependency  mechanism ensures that forces are only computed when necessary, e.g. components with zero weight are not evaluated, and components that have already been computed at an inner level are reused. 
}

\subsection{Spatially localized ring polymer contraction}

\begin{figure}[ht]
    \centering
    \includegraphics[width=\textwidth]{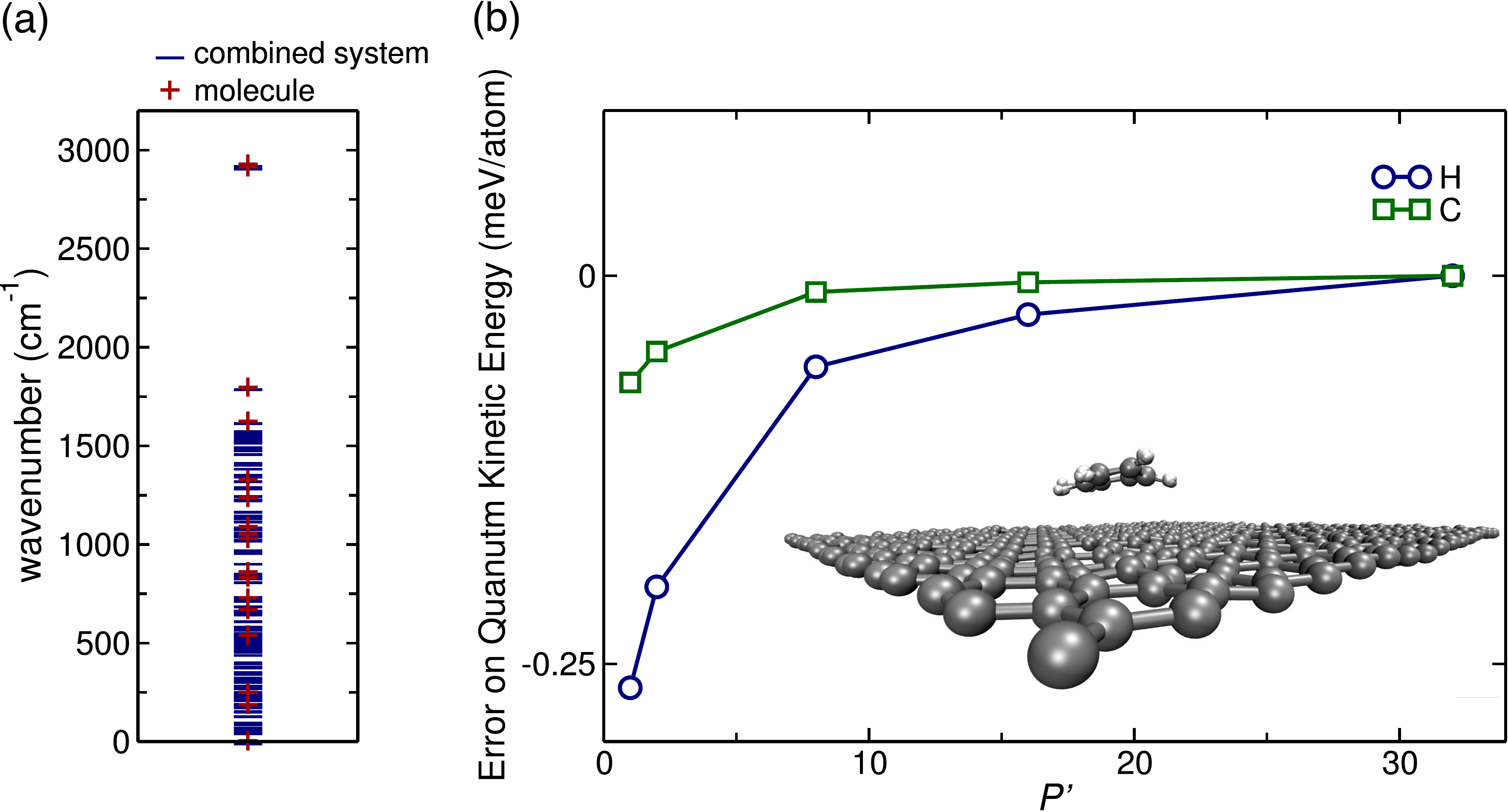}
    \caption{(a) Comparison of the vibrational frequencies of the full system (blue bars) and of the benzene molecule at the geometry it adopts on top of graphene (red crosses). (b) Convergence of the quantum kinetic energy on the adsorbate for the SL-RPC scheme \rev{with respect to the contracted number of beads $P'$ (see Eq. \ref{eq:v-slc})}.}
    \label{fig:sl-rpc}
\end{figure}

As discussed in Ref. \cite{litm+17jcp}, the spatially localized ring polymer contraction (SL-RPC) is particularly useful when dealing with molecules weakly adsorbed on surfaces and can be seen as a simplified version of the adaptive multiresolution method~\cite{Kreis2016}. 
In such cases, it is possible to seamlessly define a natural partition of the system into surface and adsorbates. In situations where one wishes to investigate the quantum mechanical behavior of
the nuclei of the adsorbate, it has been shown in Ref. \cite{litm+17jcp} that it is possible to grasp the effect of quantum statistics without having to pay the price of a full path integral simulation for the whole system. 
The full system is thus treated at a reduced number of beads using ring-polymer contraction (ideally, performing a full contraction to the centroid) and  only the adsorbate is modelled with  the full number of beads.
More specifically, the  potential is approximated by
\begin{equation}
V_P(\bq) \approx 
\frac{P}{P'}\sum_{k=1}^{P'} \left[V_\tf(\tilde{\bq}^{(k)})-V_\tm(\tilde{\bq}_\tm^{(k)})\right]
+ \sum_{k=1}^P V_\tm(\bq_\tm^{(k)}),
\label{eq:v-slc}
\end{equation}
where $P$ is the number of beads of the full ring polymer, $k$ runs over the beads, $\bq$ indicates the coordinates of the beads,
and $\tilde{\bq}^{(k)}$ refers to the coordinates of a ``contracted'' ring polymer of $P'$ beads obtained by Fourier interpolation of the full ring polymer \cite{mark-mano08jcp}. \rev{The quantities $V_\tm$ and $V_\tf$ correspond to the potential energy of the (isolated) adsorbate and the full system, respectively}.

In order to implement SL-RPC in \ipi{}, we have added functionality that allows different \rev{{\bf ForceField} sockets, called {\bf FFSocket} within \ipi{},} to deal with 
systems containing different numbers of atoms. \rev{To make it clearer, each {\bf FFSocket} receives forces and energies from an external client force code}. \ipi{} keeps track of which atoms are being treated by which clients
and, \rev{making use of the MTS ifrastructure,} combines the forces into a total force that is used to evolve all the degrees of freedom in the system.
This functionality can have much broader applications; for example, in the  future it will make implementing QM/MM schemes within \ipi~more straightforward.

Figure \ref{fig:sl-rpc} shows the results from an example that demonstrates the implementation of this method in i-PI. The figure is the result of a simulation of a benzene molecule adsorbed on a graphene
sheet, both described using the AIREBO\cite{airebo} potential as implemented in LAMMPS~\cite{plim95jcp}. As discussed in Ref.\cite{litm+17jcp}, the
validity of SL-RPC for the system at hand can be assessed by comparing the frequencies of vibration localized on the molecule when it is adsorbed
on a surface and the frequencies of vibration for this molecule in the gas phase, but at the geometry it adopts when it is adsorbed on
the surface. The error induced by the SL-RPC approximation becomes larger when the differences in the frequencies between these two situations gets larger.
This comparison is shown in Fig.\ \ref{fig:sl-rpc}(a) for the benzene-on-graphene system. Phonons were calculated using the \ipi{} code connected to LAMMPS and the same AIREBO potential.
Panel \ref{fig:sl-rpc}(b) shows the quantum kinetic energy on the atoms of the adsorbate molecule when using SL-RPC with different numbers of contracted beads.  The results from these simulations are compared to a reference simulation in which the full system was described using 32 beads. One notices that, because the frequencies of vibration of the adsorbate change minimally upon adsorption, the SL-RPC
scheme indeed yields a negligible error, well below 1 meV/atom, on the quantum kinetic energy of the adsorbate, even with full contraction to the centroid ($P'=1$).

\subsection{Ring-polymer instantons}

One of the most efficient methods for simulating quantum nuclear effects
in molecular systems
is semiclassical instanton theory \cite{Perspective,Miller1975semiclassical}.
This approach can be used to compute tunneling splittings
and to calculate the rate constant for a reaction through an energy barrier
\cite{InstReview}.
Instanton theory is based on a well defined dominant tunnelling pathway which includes corner-cutting effects.
These methods can be derived from steepest-descent approximations to
a path-integral description of the tunneling process \cite{Miller1975semiclassical,AdiabaticGreens,InstReview}.

The ring-polymer instanton method \cite{RPInst,Andersson2009Hmethane,Rommel2011locating}
uses a discrete representation for the path integrals
in a similar manner to path-integral molecular dynamics.
The procedure for obtaining the ring-polymer instanton rate is closely related to the commonly-used Eyring approach for classical transition-state theory.
In both cases, the translational, rotational and vibrational partition functions for the reactants are compared with those at the transition-state.
The major difference is that,
at low temperatures, the saddle point is no longer identified as the transition state for the reaction. Instead, it corresponds to a stretched ring polymer configuration,
known as an instanton.
\rev{This is defined as the saddle point of the ring-polymer potential, given by
\begin{equation}
    U_{P}(\bq)=\sum_{k=1}^{P} \frac{1}{2}m\omega_P^2(\bq_k-\bq_{k-1})^2+\sum_{k=1}^{P} V(\bq_k). \label{eq:instpot}
\end{equation}

The instanton action is then given by $S=\hbar \beta_P U_P(\tilde{\bq})$, where $\tilde{\bq}$ is the optimized instanton geometry.
Each replica corresponds to a different imaginary time slice between 0 and $\beta\hbar$
and together they describe a tunnelling pathway.
The energy of this pathway }
is lower than the barrier top
and as a consequence allows one to account for the speed-up of the reaction rate due to tunneling.
Tunneling becomes particularly important when reactions happen below the so-called
cross-over temperature, which is defined in the harmonic approximation as $T_c=\hbar \omega^\ddagger/(2 \pi k_B)$, where $\omega^\ddagger$
is the imaginary vibrational frequency at the classical saddle point.

Instanton rate theory is thus a generalization of transition-state theory
which includes delocalization, tunneling and zero-point energy effects.
In this way, it is closely related in formulation to RPMD rate theory \cite{RPInst}.
As no path-integral sampling is necessary, the efficiency is expected to be much higher than RPMD rate theory,
and in typical calculations, its bottleneck is the evaluation of a number of ab initio Hessians.
It is important to note, however, that the instanton approach is not applicable to liquid systems as a harmonic approximation is not valid in this case.  There are thus important cases when RPMD should be used to obtain the rate instead.

A related ring-polymer instanton approach can be used to compute tunnelling splittings for degenerate rearrangements \cite{tunnel}.
The methodology is similar to the rate calculation, except that the tunnelling splitting is defined in the $T\rightarrow0$ limit
where the instanton is not a saddle point, but a minimum.

This ring-polymer instanton method
has been used to compute tunneling rates for a wide range of reactions
from gas-phase collisions to enzyme-catalysed proton transfers
\cite{HCH4,Rommel2012enzyme,Asgeirsson2018instanton}.
It has also been applied to predict tunneling splittings in a number of molecules and molecular clusters
\cite{Milnikov2001,hexamerprism,Cvitas2018instanton}.

The ring-polymer instanton method that has been implemented in i-PI uses the newly available {\bf Motion} class. It accepts two modes, namely ``rates" (for the calculation of reaction rates) and
``splitting" (for the calculation of tunneling splittings). In both of these modes, only half of the ring polymer is considered when computing the instanton geometry, and ring polymer symmetry is used to calculate all quantities related to the instanton \cite{Andersson2009Hmethane,Rommel2011locating}. The optimization algorithms for the instanton geometry that we have implemented are the quasi-Newton Nichols algorithm \cite{Nichols1990mep} and the Newton-Raphson
algorithm for rate calculations, and l-BFGS \cite{Liu1989lBFGS} for tunneling splitting calculations. The calculation of rates with the first two algorithms requires the Hessian of all replicas to be calculated at the very first step. Within one instanton calculation, these
Hessian matrices are then updated using the Powell algorithm. \cite{Fletcher}

In practice, the most efficient way to perform an instanton calculation is to start at a temperature just slightly below $T_c$, which can be much higher than the target temperature of interest.  The advantage of starting at a higher temperatures is that the instanton geometry will require a reduced amount of replicas (beads) to achieve convergence. An initial guess to position these replicas can be provided
by hand (using any interpolation scheme the user likes) or calculated automatically if the associated classical transition state geometry is known and if the Hessian matrix of this state is provided.
After converging this first instanton geometry, the temperature can be lowered and a new calculation, with, if necessary, an increased number of beads, can be restarted from the previous instanton geometry.
As the number of beads will need to be increased progressively as one lowers the temperature of the instanton simulation, it is desirable not to recalculate the Hessian matrices for each new replica, as it is an effort that may
\rev{become very} expensive.
\rev{For example, at very low temperatures, one may need more than 2000 replicas \cite{hexamerprism} to achieve convergence. Since we wish to treat systems containing around 100 atoms (or more), 
in a naive instanton calculation just the effort of computing Hessians by finite differences could amount to more than 10$^6$ force evaluations.}
We have, therefore, derived and implemented a ring polymer
expansion of the Hessian \cite{litm-preparation}. One can thus decrease the temperature in incremental steps until the target temperature is reached, without calculating any new Hessian matrices \rev{for finding the instanton geometry.}
The instanton geometry with a converged number of beads at the target temperature does not give immediate access to the rate or tunnelling splitting values. \rev{For the final calculation of the instanton rates, for example, the Hessians of all replicas are still needed. In our tests, with our ring polymer expansion, we could obtain converged results for this rate at low temperatures with 1/4 the amount of beads.} We provide post-processing routines within
the i-PI framework that \rev{calculate rates and tunneling splittings.}

\begin{table}[ht]
    \centering
    \begin{tabular}{c|c}
    $T$ (K)& $\kappa_{\text{tun}}$ \\
    \hline
    \hline
       150  & 94992 \\ %
       200  & 190.3 \\ %
       250  & 18.94 \\ %
       300  & 9.75 \\  %
    \end{tabular}
    \caption{Tunneling enhancement factors for the CH$_4$+H $\rightarrow$ CH$_3$+H$_2$ reaction with CBE potential at different temperatures.}
    \label{tab:instanton}
\end{table}

\begin{figure}
    \centering
    \includegraphics[width=0.5\textwidth]{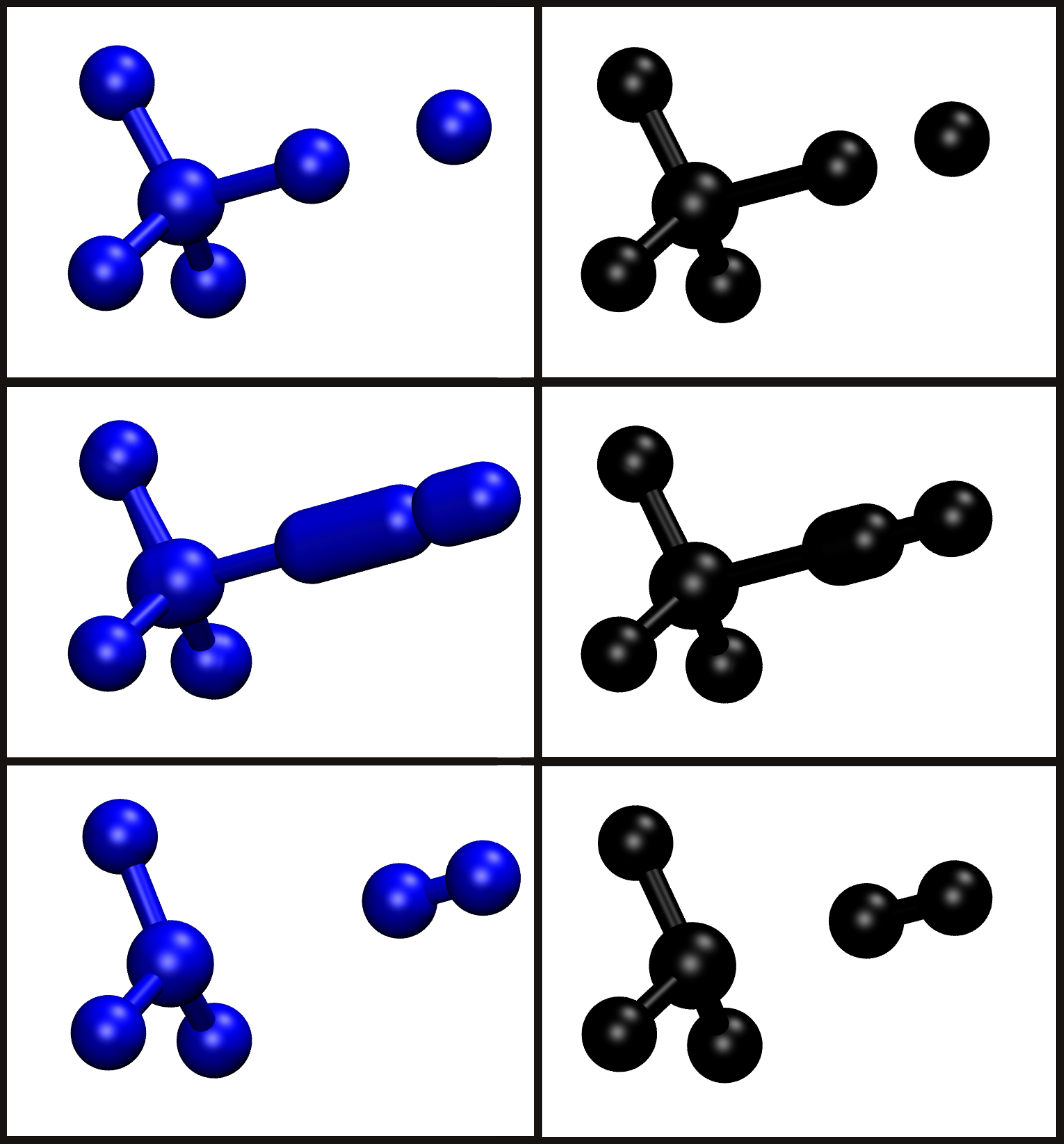}
    \caption{CH$_4$+H instanton geometries at 250 K (left) and 300 K (right) for three situations. Top: $\tau=0$; Middle: All imaginary time slices (instanton geometry); Bottom: $\tau=\beta \hbar/2$.  }
    \label{fig:instgeo}
\end{figure}

\begin{figure}
    \centering
    \includegraphics[width=0.5\textwidth]{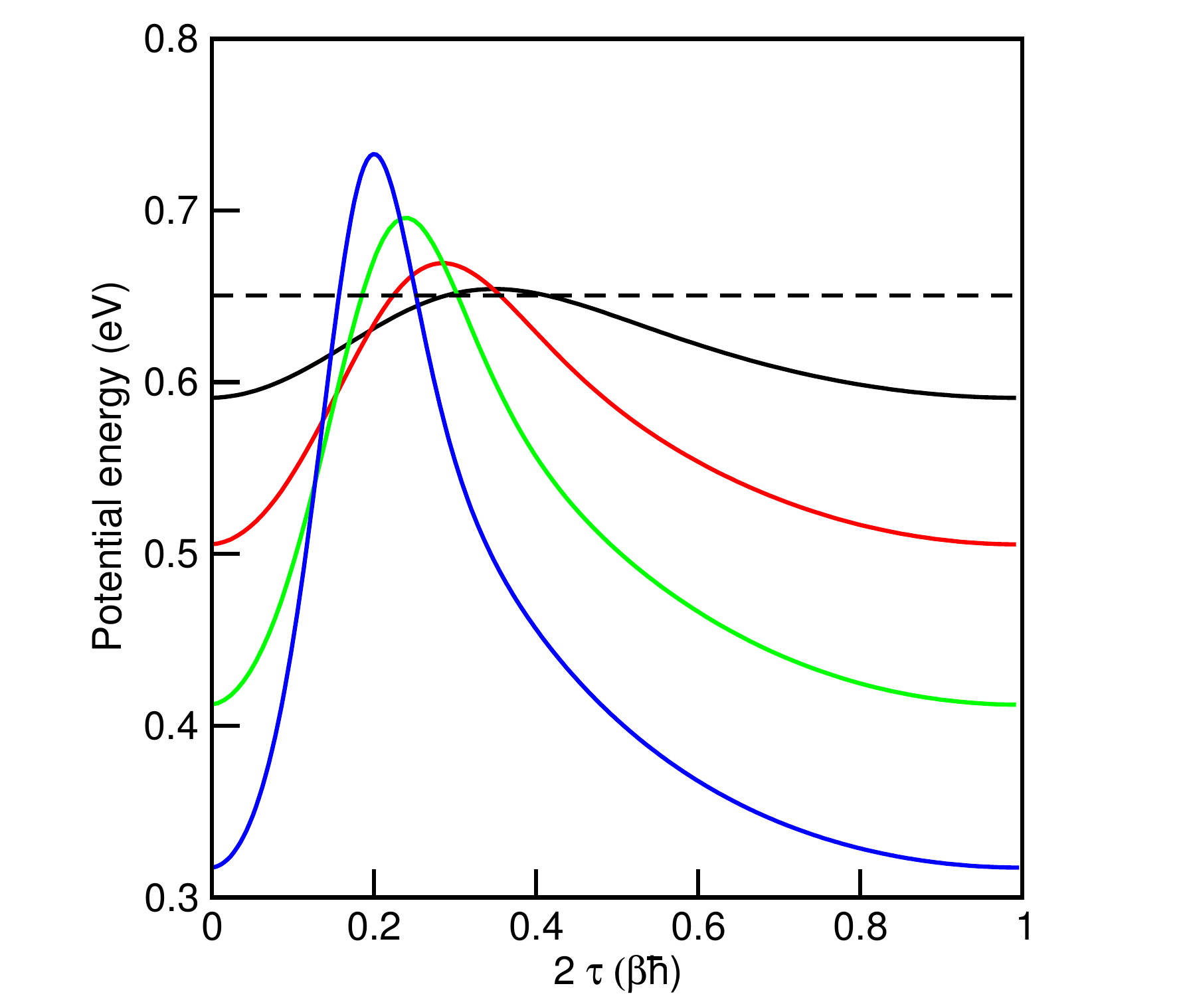}
    \caption{CH$_4$+H instanton potential energies at different imaginary-time slices at 300 K (black), 250 K (red), 200 K (green) and 150 K (blue). The classical value is given by the dashed line.}
    \label{fig:instenergy}
\end{figure}

As an example, this implementation was used to calculate the instanton reaction rates for the CH$_4$+H $\rightarrow$ CH$_3$+H$_2$ reaction at two different target temperatures, using the CBE potential \cite{cbe-potential} which has have also been included in the the $driver$ code distributed with i-PI.
The results of the ratio between the instanton rate and the Eyring transition state theory rate, i.e. the tunneling enhancement $\kappa_{\text{tun}}$, are shown in Table \ref{tab:instanton}. These are all within 0.6\% of the values obtained using the completely independent implementation of this method that was employed in Ref. \cite{MUSTreview}. 
The instanton geometry \rev{(see Eq. \ref{eq:instpot})}, and bead-geometries corresponding to imaginary time \rev{slices between} $\tau=0$ and  $\tau = \beta\hbar /2$  for 250K and 300 K are shown in Figure \ref{fig:instgeo},
while the potential energies associated with these different imaginary time slices at all different temperatures are shown in Fig. \ref{fig:instenergy}. The comparison to the classical potential energy in Fig. \ref{fig:instenergy}
highlights that at different imaginary time slices, the instanton assumes energies that are lower than the classical barrier.

\subsection{PLUMED Interface: Metadynamics of the Zundel Cation}

In a collaboration with the developers of the PLUMED code~\cite{plumed} -- a plugin for free-energy calculations that can be combined with several molecular modelling packages -- a native interface was developed that allows collective variable values and biases from PLUMED to be directly evaluated from \ipi{}.  
This  interface provides an example of a tighter integration between the client code and i-PI. Tight integration is important in this case because, in order to generate the history-dependent bias that is used e.g.\,for metadynamics calculations~\cite{laio-parr02pnas,bard+08prl}, there must be a synchronization between i-PI and the driver. This synchronization breaks the strict encapsulation that underlies the socket-based communication protocol. 
Consequently, the PLUMED interface is implemented using a {\bf ForceField} base class, and can be used both as a force provider in a {\bf System} class, or as a bias in an {\bf Ensemble} class. Crucially, it also contains options that specify the PLUMED working directory, a counter to hold the number of metadynamics steps that have been performed and a method to advance metadynamics by deploying hills at the interval specified in the PLUMED input. This solution allows calculations to be interrupted and restarted in a consistent state, as long as PLUMED's own restart files are not tampered with. 

\begin{figure}
\centering\includegraphics[width=0.48\textwidth]{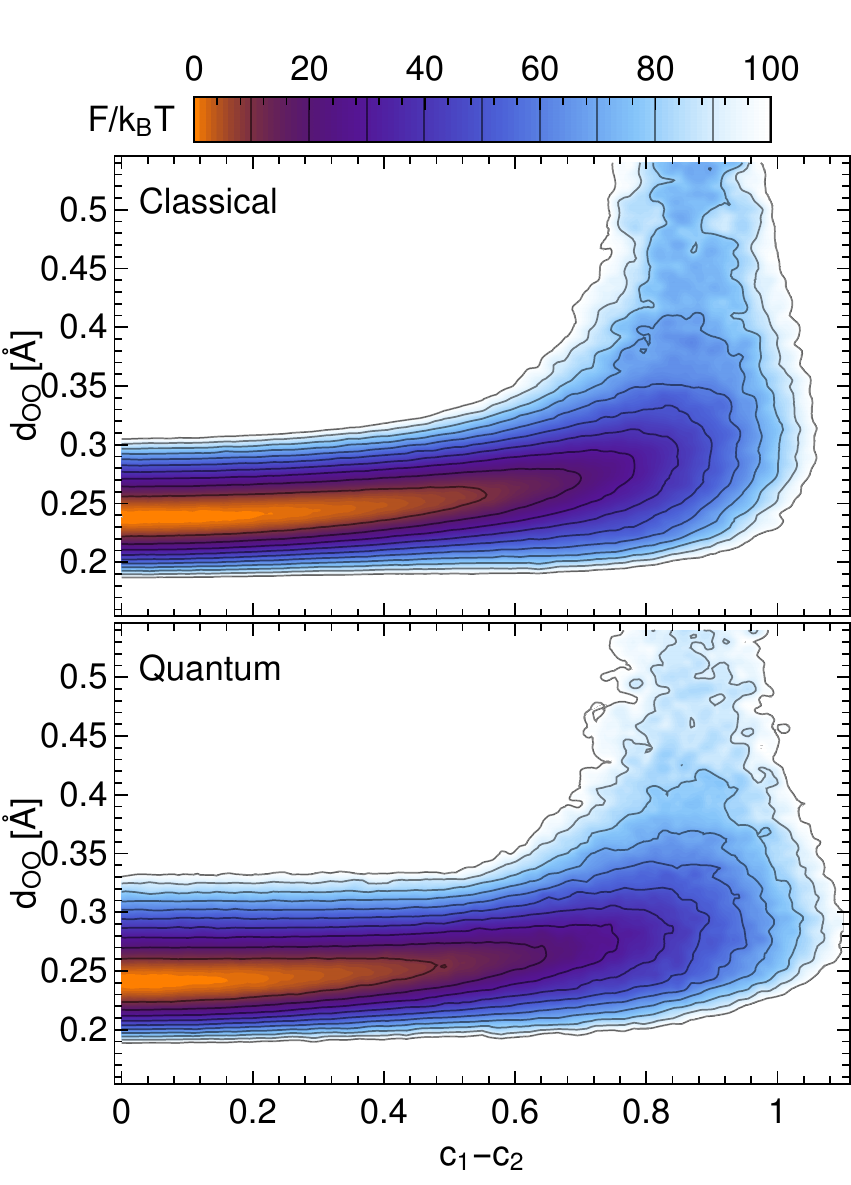}
\caption{
Free-energy surface for a classical (top) and quantum (bottom) simulation of the Zundel cation at 150~K. Lines indicate intervals of 10 $k_B T$.
$d_\text{OO}$ indicates the distance between the O atoms, and $c_1$ and $c_2$ their hydrogen coordination, computed by summing the value of a sigmoid function of the distance between the protons each of the two O atoms. 
\label{fig:zundel-fes}
}
\end{figure}

To demonstrate how this interface works, a metadynamics simulation of the Zundel cation \ce{H5O2+} was performed using a potential energy surface that was fitted to high-end quantum chemistry calculations~\cite{huan+05jcp}. 
We computed the free-energy surface as a function of the distance between the O atoms $d_\text{OO}$, and the difference in their hydrogen coordination numbers $c_1-c_2$, which makes it possible to distinguish \ce{H5O2+} states (with $c_1 \approx c_2$) from decomposed \ce{H2O+H3O+} configurations.
When the O atoms are close to their equilibrium distance of $\approx 2.5$\AA{} the proton is delocalized in a Zundel-like state. When the O atom are pulled apart, the complex decomposes in a neutral water molecule and an Eigen cation, and there is a large barrier to transfer the proton at fixed $d_\text{OO}$. 
We performed calculations at 150~K, using both a classical trajectory and one with PIMD, using 64 beads, and a  bias that was deployed on the centroid. 
\rev{Fig.~\ref{fig:zundel-fes} compares  classical and quantum results for the free energy of the centroid, computed along the proton-transfer coordinate and the O--O distance. 
Effects are small but noticeable. The  fluctuations of the O--O distance are enhanced by quantum effects. However, perhaps counter-intuitively, nuclear quantum effects tend to stabilize the symmetric shared-proton state, and to destabilize  the dissociated \ce{H2O{} + {}H3O+} state, as one can see by the faster increase of the free energy in the large $d_\text{OO}$ region in the quantum simulation.
}

\subsection{Perturbed path-integrals (PPI)}	
	
The accuracy of standard quantum estimators for energies, structural and response properties can be substantially improved by combining imaginary time path-integrals (PI) with quantum-mechanical perturbation theory~\cite{polt-tkat16cs, polt+18jcp}. Within this method, the atomic forces are used to improve the partition function of the ring-polymer, $Z_{\rm PI}$, as:
\begin{equation}
Z_{\rm PPI} = Z_{\rm PI} \times \exp\left\lbrace-\frac{\hbar^2\beta^3}{24 P^3}\sum_{s=1}^P\sum_{n=1}^N\frac{1}{m_n}\left\langle \vec{f}^{\,(s)}_n\,^2\right\rangle_{\rm PI}\right\rbrace\,.
\label{eq:ppi}
\end{equation}
Here $P$ is the number of beads used for PIMD simulations, $N$ is the number of particles in the system, $\beta$ is the inverse temperature,  $m_n$ is the mass of particle $n$, and $\vec{f}^{\,(s)}_n$ is the force acting on particle $n$ within bead $s$.  All the required PPI estimators can be derived from $Z_{\rm PPI}$ in the standard way~\cite{polt-tkat16cs}.
	
The correction term on the RHS of Eq.~\eqref{eq:ppi} can be computed using quantities taken from a standard PIMD run, using post-processing tools that are included in this release of i-PI. Hence, the PPI approach does not increase the cost of a PIMD calculation and does not require further implementation effort.
	
\begin{figure}[hbtp]
\centering
\includegraphics[width=0.6\textwidth]{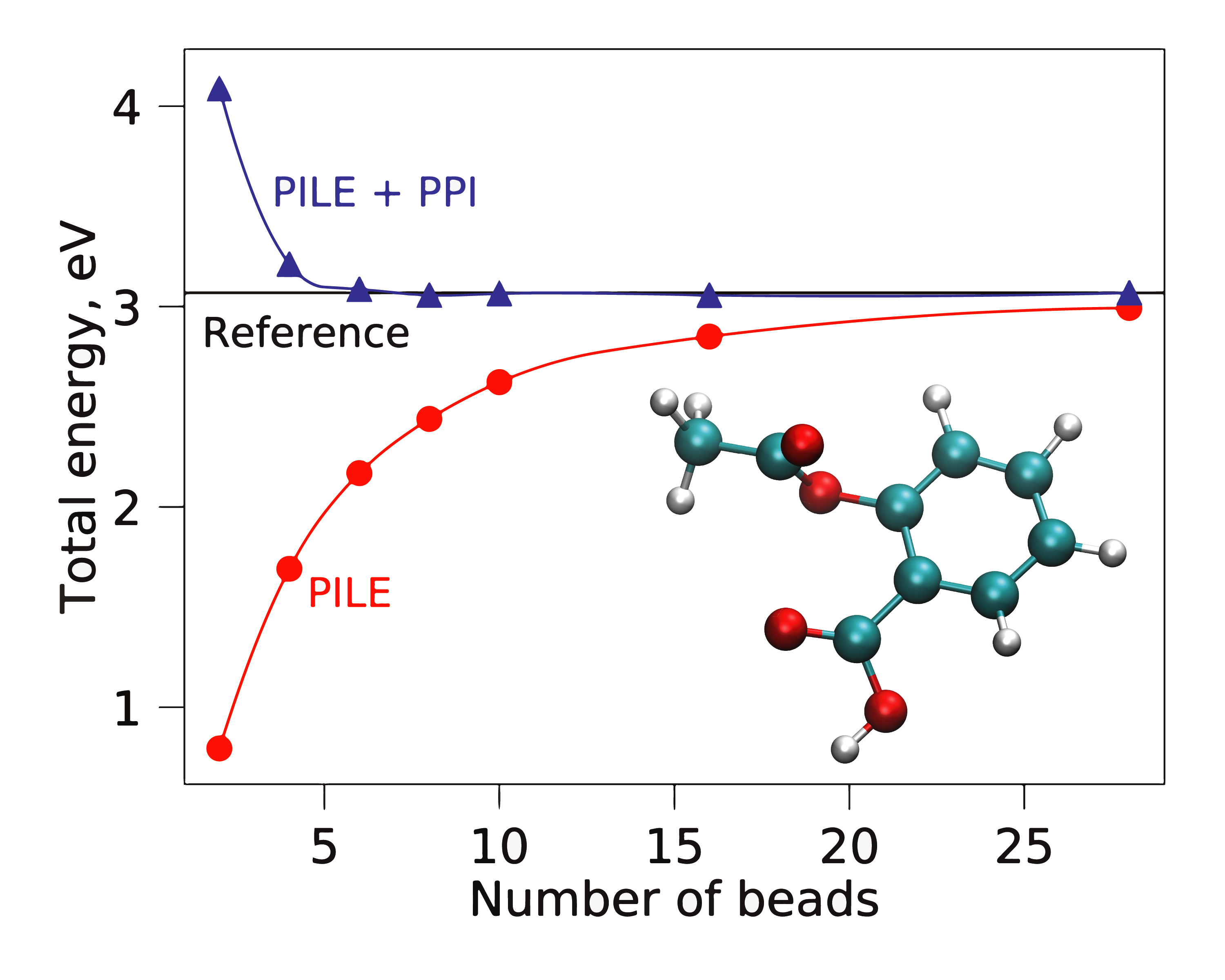}
\caption{The NQE contribution to the total energy of an aspirin molecule in the gas phase computed using different numbers of beads. The second-order PIMD simulations with Langevin equation thermostat (PILE) have been done at 300 K with a time step of 0.1~fs. The total energy is computed with (blue) and without (red) the PPI correction.}
\label{fig:ppi}
\end{figure}

The efficiency of PPI estimators is demonstrated for the  total energy of a gas-phase aspirin molecule at ambient conditions in figure \ref{fig:ppi}. The molecule is described by employing a recently developed sGDML force-field~\cite{sGDML}, which can be easily interfaced to the i-PI code. One can see that the nuclear quantum effects (NQE) contribution to the total energy can be described with 0.5~\% accuracy using only six beads when employing the PPI estimator. For comparison, the conventional virial total energy estimator underestimates the quantum part of the total energy by 2.5~\% even when twenty-eight beads are used.

\section*{Conclusions}

This second release of \ipi{} makes it even easier to use more classical and quantum sampling techniques with any electronic-structure, machine-learning or empirical force field code as a driver. 
The structure of the code has been tidied up, and the description of a physical system and its sampling have been encapsulated into separate objects, which greatly simplifies the conception and implementation of replica exchange techniques. 
The modular structure of \ipi{} makes it possible to easily combine advanced sampling techniques.  For example an Hamiltonian replica-exchange simulation with open path integrals and multiple-time-step integration can be easily realized without having to write a single line of code.
The implementation of a direct communication protocol with PLUMED demonstrates an example of a tighter integration between i-PI and the driver codes, that might be beneficial when running on a high-performance computer system.
The implementation of efficient and user-friendly post-processing tools complement the core engine, simplifying the analysis of simulations.
Future developments, bugfixes, tutorials and examples will be made available on \url{http://ipi-code.org}.

\section*{Acknowledgements}

Development of \ipi{} was directly funded by the  European Center of Excellence MaX - Materials at the Hexascale - GA No. 676598 (RP). 
Additional funding was provided by the Swiss National Science Foundation, Projects 200021-159896 (VK and BC) and 175696 (JOR); the  European Union’s Horizon 2020 research and innovation programme, Grant Agreement No. 677013-HBMAP (MC, RHM and DW), No. 716142 (TS, JK and TDK) and No. 647755 (SV, JW and VVS); The National Centre of Competence in Research (NCCR) Materials Revolution: Computational Design and Discovery of Novel Materials (MARVEL) of the Swiss National Science Foundation (SNSF) (AC and CC), the National Science Foundation under Grant No. CHE-1652960 (TEM and OM), the Fund for Scientific Research - Flanders (FWO) (SV, JW and VVS). Part of the work presented here was supported by a grant from the Swiss National Supercomputing Centre (CSCS) under project ID s719 and s786 (MR, YL, MC).

\bibliographystyle{elsarticle-num}

\begin{thebibliography}{10}
\expandafter\ifx\csname url\endcsname\relax
  \def\url#1{\texttt{#1}}\fi
\expandafter\ifx\csname urlprefix\endcsname\relax\def\urlprefix{URL }\fi
\expandafter\ifx\csname href\endcsname\relax
  \def\href#1#2{#2} \def\path#1{#1}\fi

\bibitem{ceri+14cpc}
M.~Ceriotti, J.~More, D.~E. Manolopoulos, {i-PI: A Python interface for ab
  initio path integral molecular dynamics simulations}, Comp. Phys. Comm. 185
  (2014) 1019--1026.

\bibitem{mark-ceri18nrc}
T.~E. Markland, M.~Ceriotti, {Nuclear quantum effects enter the mainstream},
  Nature Reviews Chemistry 2 (2018) 0109.

\bibitem{ceri+10jcp}
M.~Ceriotti, M.~Parrinello, T.~E. Markland, D.~E. Manolopoulos, {Efficient
  stochastic thermostatting of path integral molecular dynamics.}, J. Chem.
  Phys. 133 (2010) 124104.

\bibitem{mark-mano08jcp}
T.~E. Markland, D.~E. Manolopoulos, {An efficient ring polymer contraction
  scheme for imaginary time path integral simulations.}, J. Chem. Phys. 129
  (2008) 024105.

\bibitem{mark-mano08cpl}
T.~E. Markland, D.~E. Manolopoulos, {A refined ring polymer contraction scheme
  for systems with electrostatic interactions}, Chem. Phys. Lett. 464 (2008)
  256.

\bibitem{buss+07jcp}
G.~Bussi, D.~Donadio, M.~Parrinello, {Canonical sampling through velocity
  rescaling}, J. Chem. Phys. 126 (2007) 14101.

\bibitem{ceri+09prl}
M.~Ceriotti, G.~Bussi, M.~Parrinello, {Langevin Equation with Colored Noise for
  Constant-Temperature Molecular Dynamics Simulations}, Phys. Rev. Lett. 102
  (2009) 020601.

\bibitem{ceri+10jctc}
M.~Ceriotti, G.~Bussi, M.~Parrinello, {Colored-Noise Thermostats {\`{a}} la
  Carte}, J. Chem. Theory Comput. 6 (2010) 1170--1180.

\bibitem{ceri+09prl2}
M.~Ceriotti, G.~Bussi, M.~Parrinello, {Nuclear Quantum Effects in Solids Using
  a Colored-Noise Thermostat}, Phys. Rev. Lett. 103 (2009) 30603.

\bibitem{ceri-parr10pcs}
M.~Ceriotti, M.~Parrinello, {The $\delta$-thermostat: selective normal-modes
  excitation by colored-noise Langevin dynamics}, Procedia Computer Science 1
  (2010) 1607--1614.

\bibitem{gle4md}
M.~Ceriotti, {GLE4MD}, http://gle4md.org (2010).

\bibitem{ceri+11jcp}
M.~Ceriotti, D.~E. Manolopoulos, M.~Parrinello, {Accelerating the convergence
  of path integral dynamics with a generalized Langevin equation.}, J. Chem.
  Phys. 134 (2011) 84104.

\bibitem{ceri-mano12prl}
M.~Ceriotti, D.~E. Manolopoulos, {Efficient First-Principles Calculation of the
  Quantum Kinetic Energy and Momentum Distribution of Nuclei}, Phys. Rev. Lett.
  109 (2012) 100604.

\bibitem{yama05jcp}
T.~M. Yamamoto, {Path-integral virial estimator based on the scaling of
  fluctuation coordinates: Application to quantum clusters with fourth-order
  propagators}, J. Chem. Phys. 123 (2005) 104101.

\bibitem{ceri-mark13jcp}
M.~Ceriotti, T.~E. Markland, {Efficient methods and practical guidelines for
  simulating isotope effects.}, J. Chem. Phys. 138 (2013) 014112.

\bibitem{lin+10prl}
L.~Lin, J.~A. Morrone, R.~Car, M.~Parrinello, {Displaced Path Integral
  Formulation for the Momentum Distribution of Quantum Particles}, Phys. Rev.
  Lett. 105 (2010) 110602.

\bibitem{crai-mano04jcp}
I.~R. Craig, D.~E. Manolopoulos, {Quantum statistics and classical mechanics:
  Real time correlation functions from ring polymer molecular dynamics}, J.
  Chem. Phys. 121 (2004) 3368.

\bibitem{habe+13arpc}
S.~Habershon, D.~E. Manolopoulos, T.~E. Markland, T.~F. Miller, {Ring-polymer
  molecular dynamics: quantum effects in chemical dynamics from classical
  trajectories in an extended phase space.}, Annual review of physical
  chemistry 64 (2013) 387--413.

\bibitem{cao-voth93jcp}
J.~Cao, G.~A. Voth, {A new perspective on quantum time correlation functions},
  J. Chem. Phys. 99 (1993) 10070--10073.

\bibitem{cao-voth94jcp}
J.~Cao, G.~A. Voth, {The formulation of quantum statistical mechanics based on
  the Feynman path centroid density. IV. Algorithms for centroid molecular
  dynamics}, J. Chem. Phys. 101 (1994) 6168--6183.

\bibitem{ceri+12prsa}
M.~Ceriotti, G.~A.~R. Brain, O.~Riordan, D.~E. Manolopoulos, {The inefficiency
  of re-weighted sampling and the curse of system size in high order path
  integration}, Proceedings of the Royal Society A: Mathematical, Physical and
  Engineering Sciences 468 (2011) 2--17.

\bibitem{jang-voth01jcp}
S.~S. Jang, G.~A. Voth, {Applications of higher order composite factorization
  schemes in imaginary time path integral simulations}, J. Chem. Phys. 115
  (2001) 7832--7842.

\bibitem{kapi+16jcp2}
V.~Kapil, J.~Behler, M.~Ceriotti, {High order path integrals made easy}, J.
  Chem. Phys. 145 (2016) 234103.

\bibitem{polt-tkat16cs}
I.~Poltavsky, A.~Tkatchenko, {Modeling quantum nuclei with perturbed path
  integral molecular dynamics}, Chemical Science 7 (2016) 1368--1372.

\bibitem{kapi+18jpcb}
V.~Kapil, A.~Cuzzocrea, M.~Ceriotti, {Anisotropy of the Proton Momentum
  Distribution in Water}, J. Phys. Chem. B 122 (2018) 6048--6054.

\bibitem{liu+13jpcc}
J.~Liu, R.~S. Andino, C.~M. Miller, X.~Chen, D.~M. Wilkins, M.~Ceriotti, D.~E.
  Manolopoulos, {A Surface-Specific Isotope Effect in Mixtures of Light and
  Heavy Water}, J. Phys. Chem. C 117 (2013) 2944--2951.

\bibitem{chen+16jpcl}
B.~Cheng, J.~Behler, M.~Ceriotti, {Nuclear Quantum Effects in Water at the
  Triple Point: Using Theory as a Link Between Experiments}, J. Phys. Chem.
  Letters 7 (2016) 2210--2215.

\bibitem{chen-ceri14jcp}
B.~Cheng, M.~Ceriotti, {Direct path integral estimators for isotope
  fractionation ratios.}, J. Chem. Phys. 141 (2014) 244112.

\bibitem{litm+17jcp}
Y.~Litman, D.~Donadio, M.~Ceriotti, M.~Rossi, {Decisive role of nuclear quantum
  effects on surface mediated water dissociation at finite temperature}, J.
  Chem. Phys. 148 (2018) 102320.

\bibitem{litm-preparation}
Y.~Litman, J.~O. Richardson, T.~Kumagai, M.~Rossi, All-atom all-electron
  quantum simulations for a double-hydrogen transfer reaction, submitted
  (2018).

\bibitem{ross+16prl}
M.~Rossi, P.~Gasparotto, M.~Ceriotti, {Anharmonic and Quantum Fluctuations in
  Molecular Crystals: A First-Principles Study of the Stability of
  Paracetamol}, Phys. Rev. Lett. 117 (2016) 115702.

\bibitem{kapi+16jcp}
V.~Kapil, J.~VandeVondele, M.~Ceriotti, {Accurate molecular dynamics and
  nuclear quantum effects at low cost by multiple steps in real and imaginary
  time: Using density functional theory to accelerate wavefunction methods}, J.
  Chem. Phys. 144 (2016) 054111.

\bibitem{petr+15jcc}
R.~Petraglia, A.~Nicola{\"{i}}, M.~M.~D. Wodrich, M.~Ceriotti, C.~Corminboeuf,
  {Beyond static structures: Putting forth REMD as a tool to solve problems in
  computational organic chemistry}, J. Comp. Chem. 37 (2016) 83--92.

\bibitem{ross+14jcp}
M.~Rossi, M.~Ceriotti, D.~E. Manolopoulos, {How to remove the spurious
  resonances from ring polymer molecular dynamics.}, J. Chem. Phys. 140 (2014)
  234116.

\bibitem{ross+18jcp}
M.~Rossi, V.~Kapil, M.~Ceriotti, {Fine tuning classical and quantum molecular
  dynamics using a generalized Langevin equation}, J. Chem. Phys. 148 (2018)
  102301.

\bibitem{hija+18jcp}
M.~Hijazi, D.~M.~D. Wilkins, M.~Ceriotti, {Fast-forward Langevin dynamics with
  momentum flips}, J. Chem. Phys. 148 (2018) 184109.

\bibitem{PhysRevB.73.041105}
F.~R. Krajewski, M.~Parrinello,
  \href{https://link.aps.org/doi/10.1103/PhysRevB.73.041105}{Linear scaling
  electronic structure calculations and accurate statistical mechanics sampling
  with noisy forces}, Phys. Rev. B 73 (2006) 041105.
\newblock \href {http://dx.doi.org/10.1103/PhysRevB.73.041105}
  {\path{doi:10.1103/PhysRevB.73.041105}}.
\newline\urlprefix\url{https://link.aps.org/doi/10.1103/PhysRevB.73.041105}

\bibitem{PhysRevLett.98.066401}
T.~D. K\"uhne, M.~Krack, F.~R. Mohamed, M.~Parrinello,
  \href{https://link.aps.org/doi/10.1103/PhysRevLett.98.066401}{Efficient and
  accurate car-parrinello-like approach to born-oppenheimer molecular
  dynamics}, Phys. Rev. Lett. 98 (2007) 066401.
\newblock \href {http://dx.doi.org/10.1103/PhysRevLett.98.066401}
  {\path{doi:10.1103/PhysRevLett.98.066401}}.
\newline\urlprefix\url{https://link.aps.org/doi/10.1103/PhysRevLett.98.066401}

\bibitem{Sugita1999}
Y.~Sugita, Y.~Okamoto,
  \href{https://doi.org/10.1016/s0009-2614(99)01123-9}{Replica-exchange
  molecular dynamics method for protein folding}, Chemical Physics Letters
  314~(1-2) (1999) 141--151.
\newblock \href {http://dx.doi.org/10.1016/s0009-2614(99)01123-9}
  {\path{doi:10.1016/s0009-2614(99)01123-9}}.
\newline\urlprefix\url{https://doi.org/10.1016/s0009-2614(99)01123-9}

\bibitem{Okabe2001}
T.~Okabe, M.~Kawata, Y.~Okamoto, M.~Mikami,
  \href{https://doi.org/10.1016/s0009-2614(01)00055-0}{Replica-exchange monte
  carlo method for the isobaric{\textendash}isothermal ensemble}, Chemical
  Physics Letters 335~(5-6) (2001) 435--439.
\newblock \href {http://dx.doi.org/10.1016/s0009-2614(01)00055-0}
  {\path{doi:10.1016/s0009-2614(01)00055-0}}.
\newline\urlprefix\url{https://doi.org/10.1016/s0009-2614(01)00055-0}

\bibitem{pian-laio07jpcb}
S.~Piana, A.~Laio, {A bias-exchange approach to protein folding.}, J. Phys.
  Chem.. B 111 (2007) 4553--4559.

\bibitem{Sugita2000}
Y.~Sugita, A.~Kitao, Y.~Okamoto,
  \href{https://doi.org/10.1063/1.1308516}{Multidimensional replica-exchange
  method for free-energy calculations}, The Journal of Chemical Physics
  113~(15) (2000) 6042--6051.
\newblock \href {http://dx.doi.org/10.1063/1.1308516}
  {\path{doi:10.1063/1.1308516}}.
\newline\urlprefix\url{https://doi.org/10.1063/1.1308516}

\bibitem{habe+09jcp}
S.~Habershon, T.~E. Markland, D.~E. Manolopoulos, {Competing quantum effects in
  the dynamics of a flexible water model.}, J. Chem. Phys. 131 (2009) 24501.

\bibitem{plim95jcp}
S.~Plimpton, {Fast Parallel Algorithms for Short-Range Molecular Dynamics}, J.
  Comp. Phys. 117 (1995) 1--19.

\bibitem{buss+09jcp}
G.~Bussi, T.~Zykova-Timan, M.~Parrinello, {Isothermal-isobaric molecular
  dynamics using stochastic velocity rescaling}, J. Chem. Phys. 130 (2009)
  074101.

\bibitem{habe-mano09jcp}
S.~Habershon, D.~E. Manolopoulos, {Zero point energy leakage in condensed phase
  dynamics: An assessment of quantum simulation methods for liquid water}, J.
  Chem. Phys. 131 (2009) 244518.

\bibitem{Kreis2016}
K.~Kreis, M.~E. Tuckerman, D.~Donadio, K.~Kremer, R.~Potestio,
  \href{https://doi.org/10.1021/acs.jctc.6b00242}{From classical to quantum and
  back: A hamiltonian scheme for adaptive multiresolution
  classical/path-integral simulations}, Journal of Chemical Theory and
  Computation 12~(7) (2016) 3030--3039.
\newblock \href {http://dx.doi.org/10.1021/acs.jctc.6b00242}
  {\path{doi:10.1021/acs.jctc.6b00242}}.
\newline\urlprefix\url{https://doi.org/10.1021/acs.jctc.6b00242}

\bibitem{airebo}
S.~J. Stuart, A.~B. Tutein, J.~A. Harrison, A reactive potential for
  hydrocarbons with intermolecular interactions, The Journal of Chemical
  Physics 112~(14) (2000) 6472--6486.
\newblock \href {http://dx.doi.org/10.1063/1.481208}
  {\path{doi:10.1063/1.481208}}.

\bibitem{Perspective}
J.~O. Richardson, Perspective: {R}ing-polymer instanton theory, J. Chem. Phys.
  148 (2018) 200901.
\newblock \href {http://dx.doi.org/10.1063/1.5028352}
  {\path{doi:10.1063/1.5028352}}.

\bibitem{Miller1975semiclassical}
W.~H. Miller, Semiclassical limit of quantum mechanical transition state theory
  for nonseparable systems, J.~Chem. Phys. 62~(5) (1975) 1899--1906.
\newblock \href {http://dx.doi.org/10.1063/1.430676}
  {\path{doi:10.1063/1.430676}}.

\bibitem{InstReview}
J.~O. Richardson, Ring-polymer instanton theory, Int. Rev. Phys. Chem. 37
  (2018) 171.
\newblock \href {http://dx.doi.org/10.1080/0144235X.2018.1472353}
  {\path{doi:10.1080/0144235X.2018.1472353}}.

\bibitem{AdiabaticGreens}
J.~O. Richardson, Derivation of instanton rate theory from first principles,
  J.~Chem. Phys. 144 (2016) 114106.
\newblock \href {http://arxiv.org/abs/1512.04292} {\path{arXiv:1512.04292}},
  \href {http://dx.doi.org/10.1063/1.4943866} {\path{doi:10.1063/1.4943866}}.

\bibitem{RPInst}
J.~O. Richardson, S.~C. Althorpe, Ring-polymer molecular dynamics rate-theory
  in the deep-tunneling regime: {Connection} with semiclassical instanton
  theory, J.~Chem. Phys. 131 (2009) 214106.
\newblock \href {http://dx.doi.org/10.1063/1.3267318}
  {\path{doi:10.1063/1.3267318}}.

\bibitem{Andersson2009Hmethane}
S.~Andersson, G.~Nyman, A.~Arnaldsson, U.~Manthe, H.~J{\'o}nsson, Comparison of
  quantum dynamics and quantum transition state theory estimates of the {H +
  CH$_4$} reaction rate, Journal of Physical Chemistry A 113~(16) (2009)
  4468--4478.
\newblock \href {http://dx.doi.org/10.1021/jp811070w}
  {\path{doi:10.1021/jp811070w}}.

\bibitem{Rommel2011locating}
J.~B. Rommel, T.~P.~M. Goumans, J.~K\"astner, Locating instantons in many
  degrees of freedom, Journal of Chemical Theory and Computation 7 (2011)
  690--698.
\newblock \href {http://dx.doi.org/10.1021/ct100658y}
  {\path{doi:10.1021/ct100658y}}.

\bibitem{tunnel}
J.~O. Richardson, S.~C. Althorpe, Ring-polymer instanton method for calculating
  tunneling splittings, J.~Chem. Phys. 134 (2011) 054109.
\newblock \href {http://dx.doi.org/10.1063/1.3530589}
  {\path{doi:10.1063/1.3530589}}.

\bibitem{HCH4}
A.~N. Beyer, J.~O. Richardson, P.~J. Knowles, J.~Rommel, S.~C. Althorpe,
  Quantum tunneling rates of gas-phase reactions from on-the-fly instanton
  calculations, J.~Phys. Chem. Lett. 7 (2016) 4374--4379.
\newblock \href {http://dx.doi.org/10.1021/acs.jpclett.6b02115}
  {\path{doi:10.1021/acs.jpclett.6b02115}}.

\bibitem{Rommel2012enzyme}
J.~B. Rommel, Y.~Liu, H.-J. Werner, J.~K\"astner, Role of tunneling in the
  enzyme glutamate mutase, J.~Phys. Chem.~B 116~(46) (2012) 13682--13689.
\newblock \href {http://dx.doi.org/10.1021/jp308526t}
  {\path{doi:10.1021/jp308526t}}.

\bibitem{Asgeirsson2018instanton}
V.~{\'A}sgeirsson, A.~Arnaldsson, H.~J{\'o}nsson, Efficient evaluation of atom
  tunneling combined with electronic structure calculations, J.~Chem. Phys.
  148~(10) (2018) 102334.
\newblock \href {http://dx.doi.org/10.1063/1.5007180}
  {\path{doi:10.1063/1.5007180}}.

\bibitem{Milnikov2001}
G.~V. Mil'nikov, H.~Nakamura, Practical implementation of the instanton theory
  for the ground-state tunneling splitting, J.~Chem. Phys. 115~(15) (2001)
  6881--6897.
\newblock \href {http://dx.doi.org/10.1063/1.1406532}
  {\path{doi:10.1063/1.1406532}}.

\bibitem{hexamerprism}
J.~O. Richardson, C.~P{\'e}rez, S.~Lobsiger, A.~A. Reid, B.~Temelso, G.~C.
  Shields, Z.~Kisiel, D.~J. Wales, B.~H. Pate, S.~C. Althorpe, Concerted
  hydrogen-bond breaking by quantum tunneling in the water hexamer prism,
  Science 351 (2016) 1310--1313.
\newblock \href {http://dx.doi.org/10.1126/science.aae0012}
  {\path{doi:10.1126/science.aae0012}}.

\bibitem{Cvitas2018instanton}
M.~T. Cvitas, Quadratic string method for locating instantons in tunneling
  splitting calculations, J. Chem. Theory Comput. 14 (2018) 1487--1500.
\newblock \href {http://dx.doi.org/10.1021/acs.jctc.7b00881}
  {\path{doi:10.1021/acs.jctc.7b00881}}.

\bibitem{Nichols1990mep}
J.~Nichols, H.~Taylor, P.~Schmidt, J.~Simons, Walking on potential energy
  surfaces, J.~Chem. Phys. 92~(1) (1990) 340--346.
\newblock \href {http://dx.doi.org/10.1063/1.458435}
  {\path{doi:10.1063/1.458435}}.

\bibitem{Liu1989lBFGS}
D.~C. Liu, J.~Nocedal, On the limited memory {BFGS} method for large scale
  optimization, Math. Program. 45 (1989) 503--528.

\bibitem{Fletcher}
R.~Fletcher, Practical Methods of Optimization, 2nd Edition, John Wiley and
  Sons, Chichester, 1987.

\bibitem{cbe-potential}
J.~C. Corchado, J.~L. Bravo, J.~Espinosa-Garcia,
  \href{https://aip.scitation.org/doi/abs/10.1063/1.3132223}{The hydrogen
  abstraction reaction h+ch4. i. new analytical potential energy surface based
  on fitting to ab initio calculations}, The Journal of Chemical Physics
  130~(18) (2009) 184314.
\newblock \href
  {http://arxiv.org/abs/https://aip.scitation.org/doi/pdf/10.1063/1.3132223}
  {\path{arXiv:https://aip.scitation.org/doi/pdf/10.1063/1.3132223}}, \href
  {http://dx.doi.org/10.1063/1.3132223} {\path{doi:10.1063/1.3132223}}.
\newline\urlprefix\url{https://aip.scitation.org/doi/abs/10.1063/1.3132223}

\bibitem{MUSTreview}
K.~Karandashev, Z.-H. Xu, M.~Meuwly, J.~Van{\'\i}{\v{c}}ek, J.~O. Richardson,
  Kinetic isotope effects and how to describe them, Struct. Dynam. 4 (2017)
  061501.
\newblock \href {http://dx.doi.org/10.1063/1.4996339}
  {\path{doi:10.1063/1.4996339}}.

\bibitem{plumed}
M.~Bonomi, D.~Branduardi, G.~Bussi, C.~Camilloni, D.~Provasi, P.~Raiteri,
  D.~Donadio, F.~Marinelli, F.~Pietrucci, R.~A. Broglia, M.~Parrinello,
  {PLUMED: A portable plugin for free-energy calculations with molecular
  dynamics}, Comp. Phys. Comm. 180 (2009) 1961--1972.

\bibitem{laio-parr02pnas}
A.~Laio, M.~Parrinello, {Escaping free-energy minima}, Proc. Natl. Acad. Sci.
  USA 99 (2002) 12562--12566.

\bibitem{bard+08prl}
A.~Barducci, G.~Bussi, M.~Parrinello, {Well-Tempered Metadynamics: A Smoothly
  Converging and Tunable Free-Energy Method}, Phys. Rev. Lett. 100 (2008)
  20603.

\bibitem{huan+05jcp}
X.~Huang, B.~J. Braams, J.~M. Bowman, {Ab initio potential energy and dipole
  moment surfaces for {\$}H{\_}5O{\_}2{\^{}}+{\$}}, J. Chem. Phys. 122 (2005)
  44308.

\bibitem{polt+18jcp}
I.~Poltavsky, R.~A. DiStasio, A.~Tkatchenko, {Perturbed path integrals in
  imaginary time: Efficiently modeling nuclear quantum effects in molecules and
  materials}, J. Chem. Phys. 148 (2018) 102325.

\bibitem{sGDML}
S.~Chmiela, H.~E. Sauceda, K.-R. Müller, A.~Tkatchenko, Towards exact
  molecular dynamics simulations with machine-learned force fields (2018).
\newblock \href {http://arxiv.org/abs/arXiv:1802.09238}
  {\path{arXiv:arXiv:1802.09238}}.

\end{thebibliography}

\end{document}